\documentclass[useAMS,usenatbib]{mn2e}
\usepackage{graphicx}
\usepackage[latin1]{inputenc}
\usepackage{subfigure}
\usepackage{indentfirst}
\usepackage{multicol}
\usepackage{Times} 
\usepackage{url}
\usepackage{graphicx}
\usepackage{multicol}
\usepackage{epstopdf}

\title[Rings under close encounters: Chariklo vs Chiron]
{Rings under close encounters with the giant planets: Chariklo vs Chiron}
\author[R. A. N. Araujo, O. C. Winter, R. Sfair]{ R. A. N.  Araujo$^{1}$,  O. C. Winter$^1$, R. Sfair$^1$\\ 
$^{1}$UNESP - S\~ao Paulo State University, Grupo de Din\^amica Orbital e Planetologia, CEP 12516-410, Guaratinguet\'a, SP, Brazil 
}
\begin{document}

\date{}

\pagerange{\pageref{firstpage}--\pageref{lastpage}} \pubyear{2017}
\maketitle
\label{firstpage}

\begin{abstract}
In 2014, the discovery of two well-defined rings around the Centaur (10199) Chariklo were announced. This was the first time that such structures were found around a small body. In 2015, it was proposed that the Centaur (2060) Chiron may also have a ring. In a previous study, we analyzed how close encounters with giant planets would affect the rings of Chariklo. The most likely result is the survival of the rings. In the present work, we broaden our analysis to (2060) Chiron. In addition to Chariklo, Chiron is currently the only known Centaur with a presumed ring. By applying the same method as \cite{araujo2016}, we performed numerical integrations of a system composed of 729 clones of Chiron, the Sun, and the giant planets. The number of close encounters that disrupted the ring of Chiron during one half-life of the study period was computed. This number was then compared to the number of close encounters for Chariklo. We found that the probability of Chiron losing its ring due to close encounters with the giant planets is about six times higher than that for Chariklo. Our analysis showed that, unlike Chariklo, Chiron is more likely to remain in an orbit with a relatively low inclination and high eccentricity. Thus, we found that the bodies in Chiron-like orbits are less likely to retain rings than those in Chariklo-like orbits. Overall, for observational purposes, we conclude that the bigger bodies in orbits with high inclinations and low eccentricities should be prioritized.
\end{abstract}


\section{Introduction}
\label{sec_introduction}

\begin{table*} 
\begin{minipage}{12cm}
\centering
\caption{Physical parameters and heliocentric orbital elements of (10199) Chariklo and (2060) Chiron}
\end{minipage}
\begin{tabular}{c c c c c c c}
\hline
	      & Mass (kg)		&Radius (km)		&$a$ (au)		&$e$ 		&$i$ (deg)  	&Ring Orbital radii (km)	\\
\hline	  
Chariklo$^1$  & $7.986\times10^{18}$  	&$124$			&$15.74$		&$0.171$	&$23.4$		&$R_1=390.6$  , $R_2=404.8$  			\\
Chiron$^2$    & $2\times10^{18}$    	&$74$			&$13.64$		&$0.3827$	&$6.9$		&$324$  					\\
\hline
\multicolumn{7}{l}{$^1$ \cite{braga1} and \cite{araujo2016}} \\
\multicolumn{7}{l}{$^2$ Mass and Radius from \cite{nasa2014}. Orbital elements obtained from JPL's Horizons system for the Epoch}  \\
\multicolumn{7}{l}{ MJD 57664 ($a$ - semimajor axis, $e$ - eccentricity, $i$ - orbital inclination). Ring Orbital Radii from \cite{ortiz}}. \\
\end{tabular}
\label{tab_data}
 \end{table*}

A stellar occultation of (10199) Chariklo revealed the existence of an associated pair of narrow well-defined rings \citep{braga1}. This was the first detection of a ring system orbiting a minor body. The detection was confirmed in later occultations \citep{berard}. 

Chariklo is the largest body defined as a Centaur. Centaurs exist in chaotic orbits among the giant planets and frequently have close encounters with them. The lifetime of a Centaur is on the order of ten million years (\cite{tiscareno}, \cite{Hor2004a}, \cite{Hor2004b}, \cite{araujo2016}).

The existence of a ring system around a minor body in such a wild orbital environment leads to the next question: Are the rings of Chariklo stable during close encounters with the giant planets? In a previous work, \cite{araujo2016} addressed this question by performing numerical simulations with a set of 729 clones of Chariklo with similar initial orbits around the Sun given the gravitational perturbations of Jupiter, Saturn, Uranus and Neptune. 
Throughout these simulations, all the close encounters between Chariklo and each giant planet were recorded. Using a selection of the strongest close encounters, a new set of numerical simulations was made considering rings of massless particles orbiting Chariklo. 
As a result, these studies identified the cases where the encounters were disruptive, removing the ring, and those where the ring was not removed but suffered a significant change in its orbit around Chariklo. The conclusion of \cite{araujo2016} was that the ring would survive without any significant orbital change for more than $90\%$ of the clones. A similar result was later obtained by \cite{wood}.

The possible existence of a ring around (2060) Chiron, another Centaur, was suggested by \cite{ortiz}. This indication, which has not yet been confirmed, raises the question of whether the results found for Chariklo`s rings \citep{araujo2016} are also valid for Chiron`s ring system since their present orbital elements and the system configuration (i.e., the size of the Centaur and location of the rings) are different.

Therefore, in the present work, we reproduce the study of the close encounters for a set of clones of Chiron, similar to what was done for Chariklo. Then, we identify the differences in the results for Chiron and those for Chariklo caused by the differences in Chiron's ring system and and by their distincts orbital evolutions.

The structure of this paper is as follows. In Section \ref{sec_cha_chi}, we discuss the physical and orbital differences between Chariklo, Chiron and their rings. In Section \ref{subsec_method}, we explain our experimental method; in Section \ref{sec_results}, we explain our results; and in Section \ref{sec_final},  we summarize our main results giving a prospect for detecting rings aroud Centaurs.

\section{Chiron vs Chariklo}
\label{sec_cha_chi}
\begin{table}
\begin{minipage}{9cm}
\centering
\caption{Orbital distributions of the clones of Chariklo and Chiron}
\end{minipage}
\begin{tabular}{c c c}
\hline
  Chariklo$^1$					&Chiron		 				&Delta$^2$\\
\hline	  	
  $15.720\leq a  \leq 15.760$  			&$13.620\leq a 	 \leq 13.660$		        &$\Delta a = 0.005$	 		\\
  $0.1510\leq e \leq 0.1910$	    		&$0.363\leq e \leq 0.403$			&$\Delta e = 0.005$  		\\
  $23.36^{\circ}\leq i \leq 23.44^{\circ}$    	&$6.91^{\circ}\leq i \leq 6.99^{\circ}$		&$\Delta i = 0.01^{\circ}$	  		\\
\hline
\multicolumn{3}{l}{Orbital semimajor axis and $\Delta a$ in astronomical units.} \\
\multicolumn{3}{l}{$^1$ From \cite{araujo2016}} \\
\multicolumn{3}{l}{$^2$ Defined following \cite{Hor2004a}} \\
\end{tabular}
\label{tab_clones}
\end{table}

Chiron and Chariklo are dynamically classified as Centaurs. Although there is no consensus on the definition of Centaurs, it is generally well accepted that the Centaur population is 
made up of small bodies whose orbits mainly evolve in the region between Jupiter and Neptune (see the discussion in \cite{araujo2016}).

The crossing of orbits of Centaurs with the giant planets and, consequently, the close encounters experienced by them, are quite frequent. This leads to a characteristic chaotic orbital evolution of the Centaurs (see, e.g., \cite{tiscareno, bailey, Lev1997, Hor2004a, Hor2004b}). 

Chiron was the first known object of the population of Centaurs \citep{kowal79}. Chariklo was discovered later in $1997$ by the Spacewatch program\footnote{http://spacewatch.lpl.arizona.edu/discovery.html}. The two exhibit distinct orbital and physical characteristics. From Table \ref{tab_data}, we see that Chariklo is almost twice as large as Chiron and that the orbit of Chariklo is less eccentric and more inclined than the orbit of Chiron. 

For the size and mass of Chiron, we used the lower values presented in \cite{nasa2014}.
The lower values of the mass and size of Chiron were chosen to make them as different as possible from those of Chariklo. Despite being the first Centaur to be detected and the second largest Centaur in size, Chiron's actual size is uncertain. Several measurements of its radius using a variety of techniques can be found in the literature: $71$ km \citep{Groussin}, $74$ km \citep{fernandez}, $84$ km \citep{Altenhoff}, $90$ km (\cite{Lebofsky} and \cite{bus}), and $100$ km \citep{bauer}. A follow-up campaign using infrared observations by \cite{Campins} resulted in radius measurements ranging from $74$ km to $94$ km over the period from 1991 to 1994. A more recent study using a thermal model from the Herschel Space Observatory gave an equivalent radius of Chiron of $109$ km \citep{Fornasier}, but this value may be overestimated since the emissions of the body and rings may overlap and cannot be separated. 
Although there is an uncertainty of the size of Chiron and, consequently, an uncertainty of its mass value, we will later discuss that this is not a factor in the problem addressed here.

Chariklo opened the door to a new and interesting subfield of astronomy - that of ringed small bodies. The two well-defined narrow rings of Chariklo were revealed by stellar occultations \citep{braga1, berard}. The possibility that Chiron may also have rings was proposed by \cite{ortiz}. 
More details about the rings are given in Section \ref{sec_results}.

The stability of the rings of Chariklo was analysed by \cite{araujo2016}. In that work we showed that the rings are more likely to survive during the lifetime of Chariklo as a Centaur and, thus, that the Centaurs in general may experience a propitious environment for the existence of rings. 
Here, we follow a similar approach with the difference that in the present work we consider Chiron instead of Chariklo. Our goal is to study the stability of a hypotetical ring system around Chiron. The results are then contrasted to previous results obtained for Chariklo. This approach will then allow us obtain statistical statements (subject to assumptions) for the possible existence of rings around Centaurs as a general population.

\begin{figure*}
\subfigure{\includegraphics[scale=0.5]{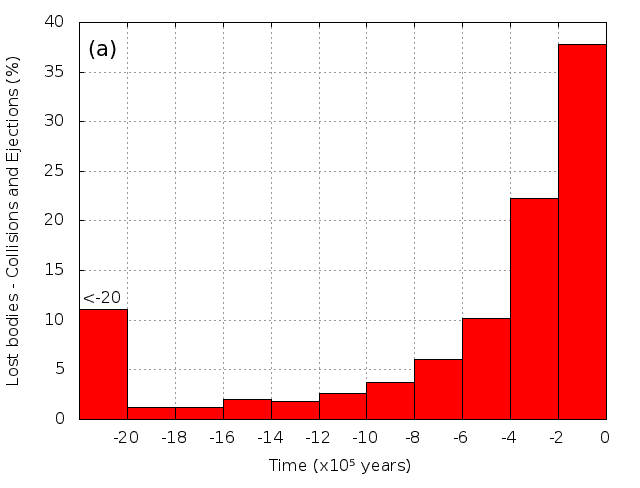}}
\subfigure{\includegraphics[scale=0.52]{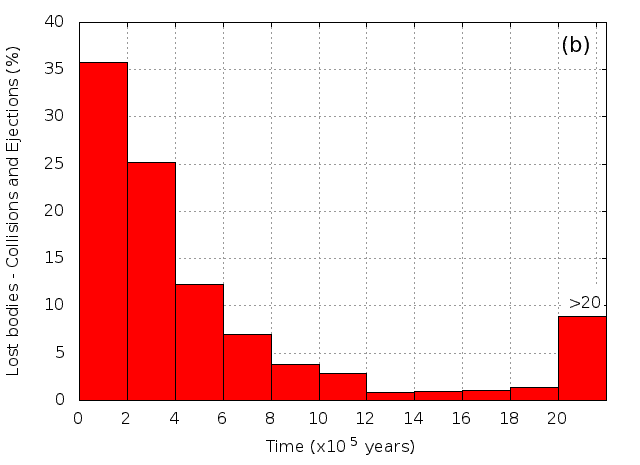}}
\caption{Histogram of the fraction of Chiron clones lost within $100$ Myr as a function of time. 
a) Backward integration. b) Forward integration. Throughout the numerical integration, the clones could be lost by ejection or 
collision with one of the giant planets or the Sun. The ejection distance was assumed to be $100$ au. A collision was assumed if the encounter distance was smaller than the body's physical radius.}
\label{fig_lifetime}
\end{figure*}

\begin{figure}
\label{fig_histo}
\centering
\subfigure{\includegraphics[scale=0.318]{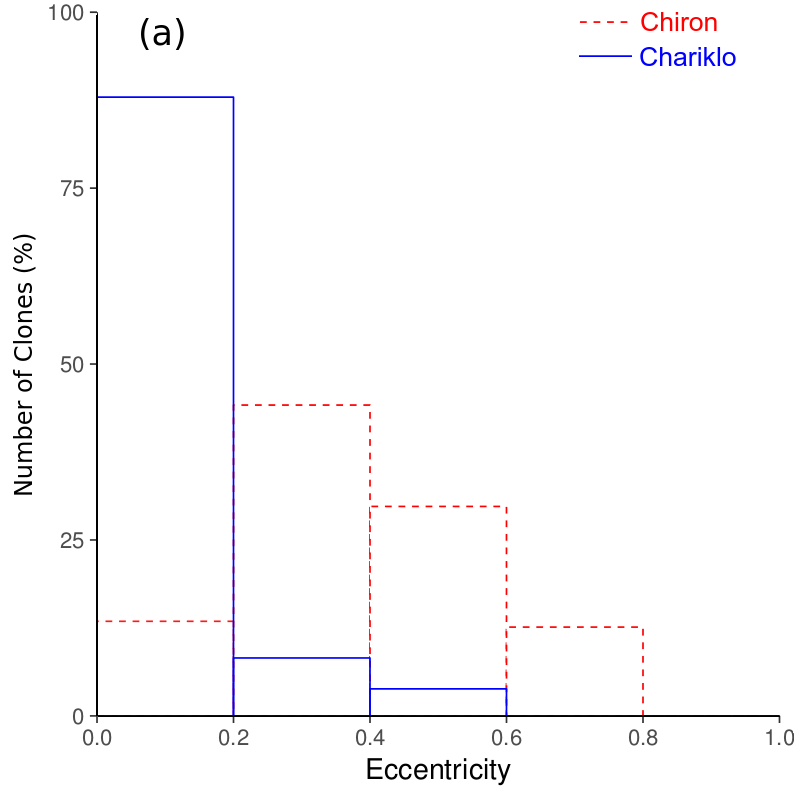}}
\subfigure{\includegraphics[scale=0.318]{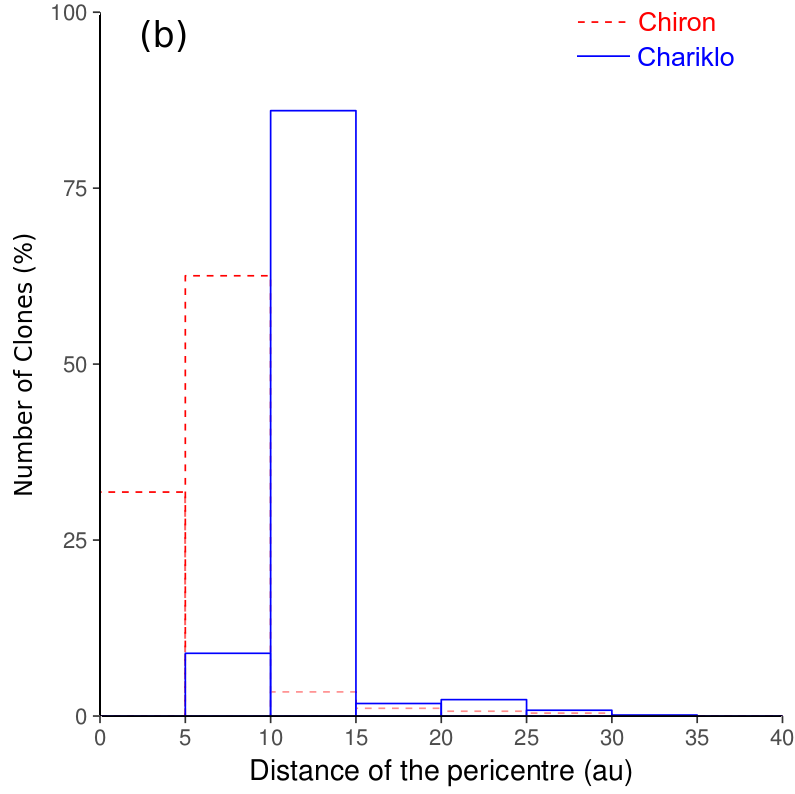}} 
\subfigure{\includegraphics[scale=0.318]{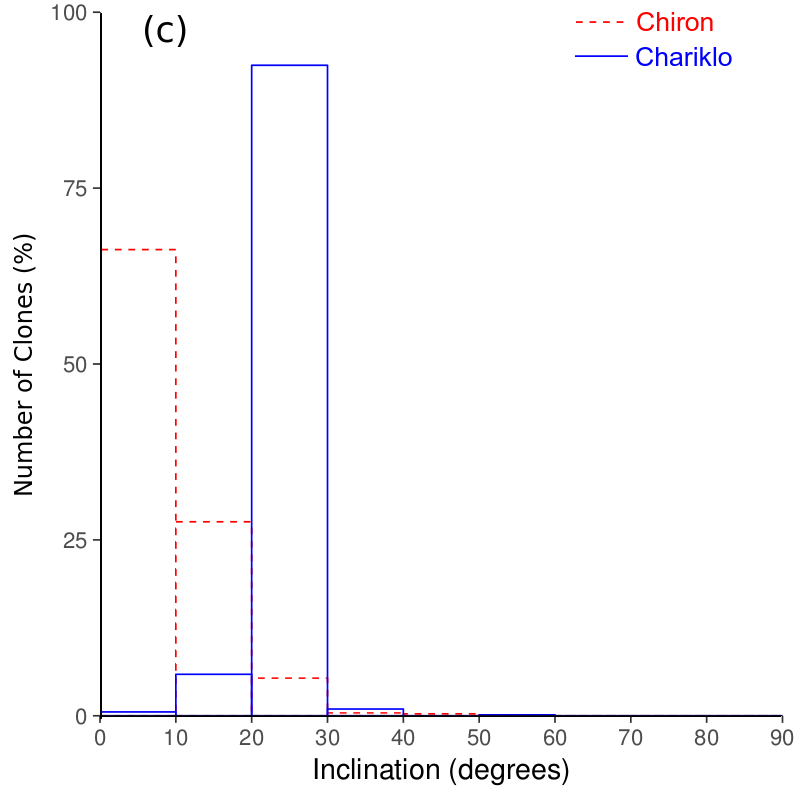}} 
\caption{Distribution of the number of the clones in terms of the eccentricities, pericentre distances and inclinations of the clones of Chiron and Chariklo. 
The histograms were made considering the temporal evolutions of these orbital elements for
all clones, finding the most likely value for each element. The data for the Chariklo case is the same as that produced in Araujo, Sfair \& Winter (2016).}
\end{figure}

\section{Method}
\label{subsec_method}
\cite{araujo2016} considered $729$ clones of Chariklo. The clones were defined from the osculating orbit of the Centaur (Table \ref{tab_data}), where the semimajor axis, the eccentricity and the orbital inclination varied within the ranges presented in Table \ref{tab_clones}. 
In an analogous fashion, $729$ clones of Chiron were also made from its osculating orbit (Table \ref{tab_data}) using the ranges presented in Table \ref{tab_clones}. They were considered nine values for the semimajor axis going from 13.620 au until 13.660 au (taken every $\Delta_a=0.005$ au), nine values for the eccentricity, going from 0.363 until 0.403 (taken every $\Delta_e=0.005$ au) and nine values for the orbital inclination going from $6.91^{\circ}$ until $6.99^{\circ}$ (taken every $0.01^{\circ}$). The combination of these values resulted in the total number of 729 clones.

The number of clones and the intervals of variarion of their osculating orbital elements for Chariklo and for Chiron were defined following \cite{Hor2004a}. The authors justify the choice of these increments by arguing that they are ``sufficiently small that the clones can be considered
as initially essentially identical to one another, yet they are large
enough to ensure that the subsequent chaotic dynamical evolution
following close planetary encounters rapidly disperse their orbits
through phase space".
The angular elements ($\Omega, \omega$ and $M$) were also obtained from JPL's Horizons system for the Epoch MJD 57664. 

The orbits of these clones were numerically integrated for a period of $100$ Myr using the adaptive time-step hybrid sympletic/Bulirsch-Stoer algorithm implemented within the MERCURY6 orbit integration package \citep{Chambers}. 
The clones were only subject to the gravitational forces of the Sun and the planets Jupiter, Saturn, Uranus and Neptune. The orbital elements of the planets were obtained from JPL's Horizons system for the same Epoch. Throughout the integration, the clones did not interact with each other, but they could collide with any of the massive bodies or could be ejected from the system. A clone was considered to have collided with a body if the encounter distance was smaller than the body's physical radius, and the ejection distance was assumed to be $100$ au \citep{araujo2016}.

All close encounters between a clone and any planet for which the separation distance was within $10~r_{td}$ were recorded. 
The tidal disruption radius $(r_{td})$ gives an approximate distance for which a given binary is expected to be disrupted by a close gravitational encounter with a more massive body \citep{philpott}. This radius is given as follows:

\begin{equation}
 r_{td}  \approx a_B\left(\frac{3M_p}{M_1+M_2}\right)^{1/3}
 \label{eq_rtd}
\end{equation}
where $M_p$ is the mass of the encountered planet, $M_1+M_2$ is the total mass of the binary and $a_B$ is the separation of the binary. 

For Chiron and Chariklo, we computed $r_{td}$ considering the external border of the ring ($324$ km for Chiron and $\approx405$ km for Chariklo) and $M_2=0$ (the rings are composed of particles). The values of the $r_{td}$ for both Chariklo and Chiron are presented in Table 3.

 The $r_{td}$ is calculated considering Chariklo and Chiron ring's systems and each one of the giant planets individually (8 values of $r_{td}$). Throughout the integrations MERCURY register all close encounters performed within a fixed value related to each one of the planets. We then set this distance to be  $10~r_{td}$ of each planet. MERCURY also provide the details of the registered close encounters, including the minimum separation between the bodies.
Thus, among all the recorded close encounters performed by Chariklo and Chiron with the giant planets within the distance of $10~r_{td}$, we selected those performed with a minimum distance within $1 ~r_{td}$, which we call disruptive encounters.

By definition, the tidal disruption radius is the distance between a binary and a more massive body within which the binary can be disrupted by tidal forces. This was confirmed by the numerical simulations of \cite{araujo2017} - see their Fig. 3. We extend this definition and state that close encounters occurring within a distance of one tidal disruption radius can also disrupt a ring. The close encounters of Chariklo with the giant planets occurring within $1~r_{td}$ were borrowed from our previous work \citep{araujo2016}. Note that, in that work, the impact parameter was used to detect the encounters instead of the minimum distance. Therefore, we also counted the number of close encounters within $1~r_{td}$ but used the minimum distance.

\begin{figure*}
\label{fig_examples}
\begin{center}
\subfigure{\includegraphics[scale=0.36]{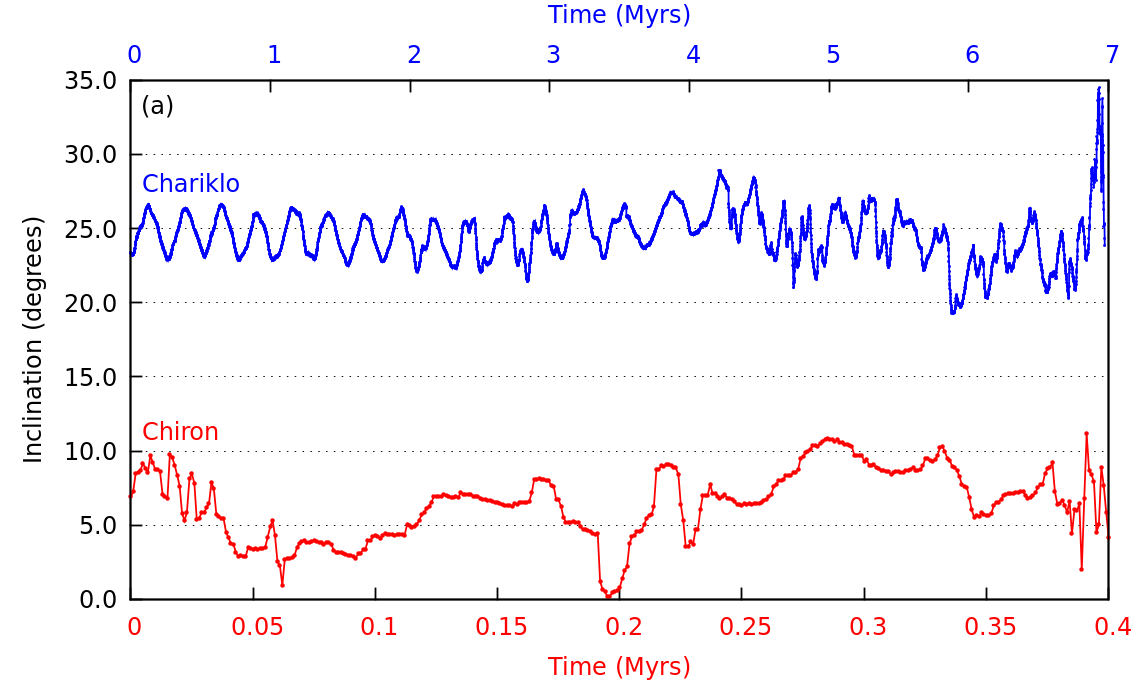}}
\subfigure{\includegraphics[scale=0.36]{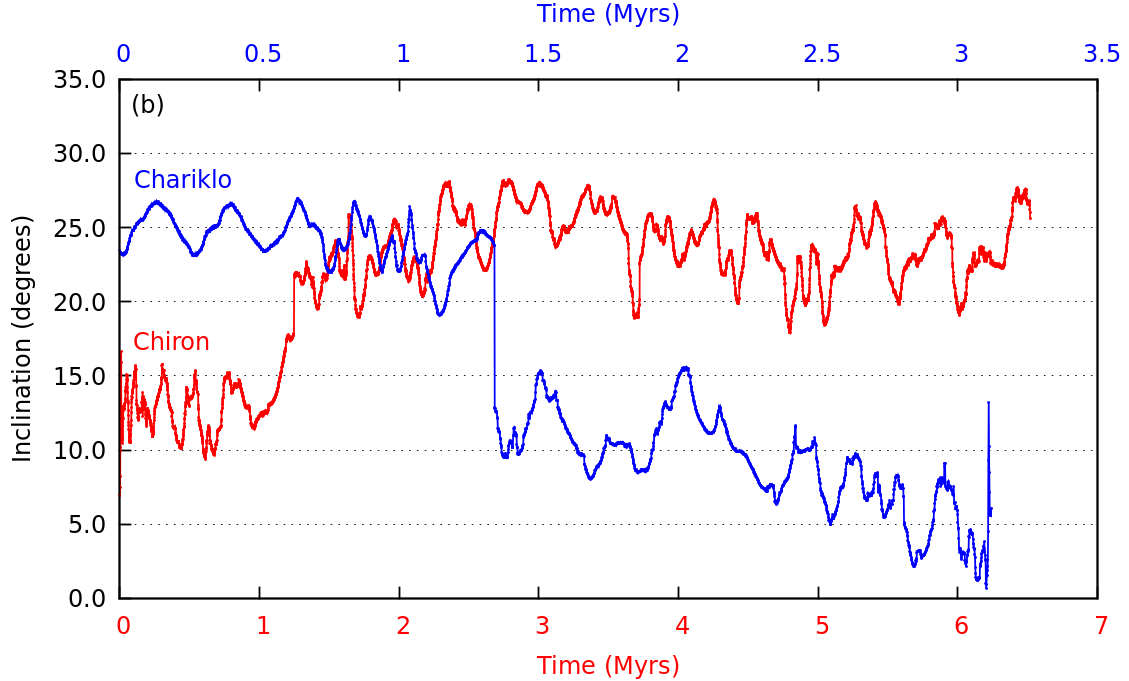}}
\caption{Examples of time evolution of orbital inclination for two statistically significant outcomes. a) the most frequent time evolution of Chariklo's orbital inclination (around $23^{\circ}$) and of Chiron's orbital inclination (around $6^{\circ}$) along their typical lifetimes. 
b) less frequent time evolution for the Chiron and Chariklo's orbital inclination . 
The lifetime of Chiron was significantly increased as it reached a relative high-inclined orbit while the lifetime of the less-inclined Chariklo was notably reduced.
}
\end{center}
\end{figure*}

\section{Results}
\label{sec_results}

\begin{table*}
\label{tab_enc}
\begin{minipage}{11.5cm}
\centering
\caption{Number of close encounters of Chariklo and Chiron performed with the giant planets within one $r_{td}$ of each body within the period of one of their respective half-lives}
\end{minipage}
 \begin{tabular}{c c c c c c c c }
\hline
\multicolumn{1}{c}{}
 &  \multicolumn{3}{c}{Chariklo}
 && \multicolumn{3}{c}{Chiron} \\ \cline{2-4} \cline{6-8}
	     	& 			&  		  &		     &	&	 	   	&	     	  &  \\
Planet		&$r_{td}$		&In Chariklo's	  &In Chriron's      &  &$r_{td}$		&In Chiron's	  &In Chariklo's  \\
	 	&(au)			&orbit		  &orbit	     &  &(au)			&orbit		  &orbit	\\	
\hline									
Jupiter	  	&$0.0025$  		&$6$		  &$32$		     &	&$0.0031 $       	&$39$    	 &$8$\\

Saturn	    	&$0.0016$ 		&$1$		  &$0$		     &  &$0.0021$	    	&$5$             &$2$\\

Uranus	     	&$0.0009$  		&$0$		  &$0$		     &	&$0.0011$	    	&$0$	         &$0$\\
	
Neptune	    	&$0.0009$ 		&$0$		  &$0$		     &	&$0.0012$	    	&$0$	         &$0$\\
\hline
		&Total			&$7$		  &$32$		     &  &			&$44$		 &$10$ \\
\hline
\end{tabular}
 \end{table*}

Accounting for how long it took for each clone to be lost (via collisions or ejections), we estimate the lifetime of Chiron as a Centaur. A histogram giving the fraction of remaining clones of Chiron as a function of time is presented in Fig. \ref{fig_lifetime}b, which shows that more than $50\%$ of the clones are lost within just $0.4$ Myr. Following \cite{Hor2004b}, we calculated the half-life of Chiron as being about $0.36$ Myr. This result is in agreement with the previous simulations made by \cite{napier2015}, who estimated a lifetime of $0.35$ Myr for Chiron.

 \cite{Hor2004a} showed that the forward and backwards half-lives of Centaurs should not always be the same. For a sample of 32 Centaurs, they found that 19 have shorter half-lives in the forward direction, while 13 have shorter half-lives in the backward direction. Although the authors concluded that for the entire data set the forward and backward half-lives diverge by a discrepan as in \cite{Hor2004a, Hor2004b} and \cite{oikawa}, we performed the same integrations for the clones of Chiron with reversed time (backward integrations).
A histogram giving the fraction of remaining clones of Chiron in the past is presented in Fig. \ref{fig_lifetime}a. By comparing these histograms, it is verified that the backward and forward half-lives of Chiron is quite symmetrical.

In our previous work \citep{araujo2016}, the study of the stability of Chariklo`s ring system was done with respect to a Centaur lifetime of 10 Myr \citep{tiscareno}. Nevertheless, revisiting our data for Chariklo, we calculated that its half-life is $7.0$ Myr. Thus, to make a fair comparison, we count the encounters of Chariklo or Chiron with the giant planets that occurred at distances $d\leq1\,r_{td}$ within $7.0$ Myr for Chariklo and within $0.36$ Myr for Chiron.

From Table 3 (columns $3$ and $6$), we see that a total of 7 disruptive encounters of the clones of Chariklo with the giant planets were recorded, while for Chiron, a total of 44 disruptive encounters were recorded. It is interesting to note that there were not registered any disruptive encounters of Chariklo and Chiron with Uranus and Neptune. The possible disruption of both ring systems involved only Jupiter and Saturn.

By comparing just the number of encounters, we see that the probability of Chiron losing its ring during one of its half-lives is approximately $44/7\approx6.3$ times higher than that for Chariklo. 
Another way to analyze these data is to compute the number of clones that experienced disruptive encounters. We found that the 7 disruptive encounters registered for Chariklo were performed by 7 distinct clones. Thus, only approximately $1\%$ of the clones of Chariklo would lose their rings due to a close gravitational encounter with the giant planets within the period of one half-life of Chariklo. For Chiron, we found that only 40 clones from 729, i.e, $5.5\%$ of its clones would experience a disruptive encounter within a time of one half-life of Chiron. 

In addition, \cite{araujo2016} showed that the past evolutions of the clones of Chariklo were symmetric with their forward evolutions. The same behaviour was found for Chiron (compare Figs. \ref{fig_lifetime}a and \ref{fig_lifetime}b). Taking this finding into account, we then estimated that approximately $2\%$ of the clones of Chariklo and approximately $11\%$ of the clones of Chiron would lose their rings due to the close encounters during their respective lifetimes as Centaurs.
Thus, by comparing the number of clones of Chariklo and Chiron that experienced disruptive encounters, we determined that the probability of Chiron losing its rings within the period of one half-life of Chiron was $40/7\approx5.7$ times higher than that for Chariklo.

This significant difference of the survivals of the rings of these two Centaurs could be attributed to the different sizes of Chiron and Chariklo and the different orbital radii of their rings as well as their different heliocentric orbital evolutions. 

Chariklo is approximately 4 times more massive than Chiron, and the external borders of their rings are of the same order of magnitude ($\approx324$ km for Chiron and $\approx405$ km for Chariklo). Consequently, as seen in Table 3, the disruption radius of Chariklo is smaller than the disruption radius of Chiron. Thus, Chariklo has to come closer to the giant planets than Chiron to have its rings disrupted by a close encounter. One may argue that this is the reason for the higher number of disruptive encounters recorded for Chiron clones.

On the other hand, Chiron has a smaller, more eccentric and less inclined orbit than Chariklo. A more eccentric orbit allows for more crossings of planetary orbits than a circular orbit, which potentially increases the number of close encounters. A less inclined orbit also increases the likelihood of close encounters, since an orbit with a lower inclination avoids long periods of time in which a small body is far outside of the planet's orbital plane where close encounters cannot occur. A small orbital semimajor axis value means that the body spend most of its period within the planetary system. Thus, the combination of the orbital elements of Chiron may also lead to the higher number of disruptive encounters found for this system than for Chariklo. We discuss this aspect as follows.

\subsection{Ring-loss probability due to Centaur mass and ring size}

To eliminate the contributions of the different heliocentric orbits and compute only the effects due to the Centaur's masses and the orbital radii of their rings, we considered both systems in the same orbit. Therefore, we considered Chiron and Chariklo in Chiron's orbit and Chiron and Chariklo in Chariklo's orbit. This was done by considering the results of the integrations of the clones of Chariklo and Chiron, selecting the disruptive encounters that each body would suffer in each of these orbits over $7.0$ Myr when in Chariklo's orbit and over $0.36$ Myr when in Chiron's orbit. That is, we computed the disruptive encounters suffered by Chariklo and Chiron if they had Chariklo's orbit, and similarly, we computed the disruptive encounters suffered by Chariklo and Chiron if they had Chiron's orbit. The results are presented in Table 3, columns $4$ and $7$. 

We see that, in both scenarios, the number of disruptive encounters experienced by Chiron is slightly higher, as expected. By comparing the total number of disruptive encounters suffered by Chariklo and Chiron in a Chariklo-like orbit (columns $3$ and $7$ of Table 3), we found that a Chariklo-like system is expected to suffer 7 disruptive close encounters with the giant planets in such an orbit, while a Chiron-like body is expected to suffer 10 such encounters. 

By comparing the total numbers of disruptive encounters suffered by Chariklo and Chiron in a Chiron-like orbit (columns $4$ and $6$ of Table 3), we estimated that a Chariklo-like system is expected to suffer 32 disruptive close encounters with the giant planets in such an orbit, while a Chiron-like body is expected to suffer 44. 

Therefore, we found that the distinct masses and orbital radii of the rings of those systems changed their chances of suffering a disruptive encounter by a factor of $10/7\approx1.43$ when in a Chariklo-like orbit, and by a factor of $44/32\approx1.38$ when in a Chiron-like orbit. 

Thus, although the masses and the orbital radii of the rings contribute to the differences in the number of disruptive encounters suffered by Chariklo and Chiron, we found that this contribution is relatively small (approximately 1.41, on average); thus, this effect alone is not enough to explain the higher probability (by a factor of $\approx 6$) of the disruption of Chiron's ring than that for Chariklo.

\subsection{Ring-loss probability due to initial heliocentric orbit}

By comparing the data in Table 3, columns 3 and 4, and columns 6 and 7, it become clear that the same ringed system would experience completely different fates when in different heliocentric orbits.
 
From columns 3 and 4 of Table 3, we see that Chariklo would experience 7 disruptive encounters when starting its integrations in a Chariklo-like orbits and would experience 32 disruptive encounters when starting its integrations in a Chiron-like orbit. On the other hand, from columns 6 and 7 of Table 3, we see that Chiron would experience 44 disruptive encounters when starting its integrations in a Chiron-like orbit and would experience only 10 disruptive encounter when starting its integrations in a Chariklo-like orbit.

The distinct initial heliocentric orbits increased the chance of Chariklo experiencing a disruptive encounter with the giant planets by a factor of $32/7\approx4.57$, while the chance of Chiron experiencing a disruptive encounter was decreased by a factor of $44/10\approx4.4$. 

On average, the heliocentric orbit led to a difference of approximately $4.5$, in terms of ring disruption. The overall probability of the contributions from the heliocentric orbital evolution and the distinct masses and orbital radii of the rings of those systems is $1.41\times4.5\approx6.3$. Thus, the heliocentric orbit of Chariklo and Chiron appears here as the main responsible for the distinct fates of the rings of these Centaurs.

\subsection{Statistics of orbital evolution}

The orbital evolution of Chiron and Chariklo were analyzed, since they showed to be major influencers on the fate of their ring systems.
To study the statistical orbital behaviours of the clones, we computed the most likely values of orbital eccentricity, the distance of the pericentre, and the inclination for each clone of Chiron and Chariklo along their respective integration times. Figs. \ref{fig_histo} present the distributions found for the whole set of clones.

From \ref{fig_histo}a, we see that more than $80\%$ of the clones of Chiron spent most of the time with $e>0.2$ and that more than $80\%$ of the clones of Chariklo spent most of the time with $e<0.2$.
Thus, Chiron is more likely to remain wandering on relatively high-eccentricity trajectories, having a higher likelihood to cross the orbits of the giant planets and thereby increasing the chance to experience a close encounter capable of removing its ring. Chariklo spends most of its lifetime wandering on relatively low eccentric trajectories. This behaviour reduces the chance that Chariklo crosses the orbits of the giant planets and has its ring system disrupted by close encounters. 

From Fig. \ref{fig_histo}b, we see that approximately $90\%$ of the clones of Chiron spent most of their time with $R_p<10$ au while approximately $90\%$ of the clones of Chariklo spent most of their time with $R_p>10$ au. Thus, most of the time, Chiron spent most of its time crossing the orbits of Saturn and Jupiter while Chariklo does not even cross the orbit of Saturn. In fact, we see that more than $30\%$ of the clones of Chiron spent most of their time in orbits with pericentre distances inside the orbit of Jupiter. This finding clearly indicates that Chiron has a higher chance of suffering from disruptive close encounters with the giant planets than Chariklo. 

From Fig. \ref{fig_histo}c, we see that more than $90\%$ of the clones of Chiron spent most of their time with $I<20^\circ$ and that more than $90\%$ of clones of Chariklo spent most of their time with $I>20^\circ$. This is a striking orbital difference between the two Centaurs. Chiron mostly wanders on low-inclination trajectories, having many more opportunities to experience very close encounters with the giant planets and to suffer perturbations strong enough to remove its ring. On the other hand, Chariklo spents long periods in trajectories with significantly higher inclinations, which drastically reduces the chances of having disruptive close encounters with the giant planets, leading to safer conditions for its rings.

Overall, when it comes to rings' stability, the evolution of the orbital inclination of the ringed Centaur plays the leading role.
The rings of a Centaur in a relatively low-inclined orbit will naturally be more perturbed when compared to a more inclined orbit.
Besides, since low-inclined orbits are more perturbed by close encounters, then, more effective they are in produce high-eccentricities and small pericentre distances. Such behaviour increase the chances of disruption of the rings of a Centaur in a initial low-inclination orbital.

Since the orbital inclination affects how perturbed an orbit is, then it will also affects the lifetime of the Centaurs.
From Fig. \ref{fig_histo}c, it has been shown that more than $90\%$ of the 729 clones of Chariklo spent most of their time with with $I>20^{\circ}$. 
The half-life for the remaining $10\%$ of clones of Chariklo that spent most of their time with $I<20^{\circ}$ was $4.1$ Myr (instead of 7 Myr, the typical half-life time of Chariklo). 
From the same figure it has been shown that more than $90\%$ of the 729 clones of Chiron spent most of their time with with $I<20^{\circ}$. 
The half-life for the remaining $10\%$ of clones of Chiron that spent most of their time with $I>20^{\circ}$ was $0.8$ Myr (instead of 0.36 Myr, the typical half-life time of Chiron).

The graphs in Figs.\ref{fig_examples}a exemplify the typical behaviour of Chiron and Chariklo ($\approx 90\%$ of the ensemble). We see that when Chariklo and Chiron maintained their typical orbital inclinations, they survived throughout the integrations along their typical half-life. 
On the other hand, the time evolution of the inclinations in Fig. \ref{fig_examples}b show an example for the less probable case for which Chiron spends most of its time in a relatively high-inclined orbit while Chariklo spends most of its time in a relatively low-inclined orbit (remaining $10\%$ of the ensemble). We see that the clones of Chiron that evolved to high inclinations survived much longer than the half-life time for the whole sample, while the survival of Chariklo was significantly reduced as its orbital inclination decreased.
All these results indicate that the lifetimes of the small bodies subject to close gravitational encounters with planets are, in fact, intrinsically related to the orbital inclinations of these bodies.

Thus, we found that the orbital evolution of Chiron is much less favourable for the existence of rings than the orbital evolution of Chariklo. This finding help to explain the results presented in Table 3. 
Overall, it becomes clear from our analysis that the orbit of a ringed Centaur determines whether the ring may experience propitious conditions for its existence rather than the configuration of the ringed system itself.

\begin{table*}
\label{tab_candidates}
\begin{minipage}{13cm}
\centering
\caption{List of small non-resonant bodies bigger than Chiron (descending order), with a distance of the pericentre 
less than $30$ au, a distance of the apocentre less than $100$ au, and with an orbital inclination greater than $20^{\circ}$}
\end{minipage}
 \begin{tabular}{c c c c c c c }
\hline
 Designation   			&$a$ 	  &$e$ 	   &Pericentre &Apocentre &$i$ 	     &Diameter \\
				&(au)     &	   &(au)       &(au)      &(degrees) & (km) \\
\hline
 2010 TY53      		&38.810   &0.458   &21.015    &56.604   &22.5     &352 \\
 2016 FX58      		&43.635   &0.391   &26.557    &60.713   &27.0     &306 \\
 2009 MF10      		&57.827   &0.527   &27.325    &88.328   &26.2     &267 \\ 
 \textbf{1997 CU26 (Chariklo)}  &15.822   &0.172   &13.099    &18.545   &23.4     &254 \\
 2007 RW10      		&30.166   &0.302   &21.070    &39.262   &36.2     &247  \\
 2013 JQ64      		&49.354   &0.543   &22.532    &76.175   &34.8     &243 \\
 \textbf{2014 NW65}      	&23.078   &0.518   &11.127    &35.030   &20.5     &232 \\
 2008 AU138     		&32.425   &0.374   &20.311    &44.539   &42.8     &202 \\
 \textbf{2000 QC243}     	&16.445   &0.200   &13.150    &19.741   &20.7     &198  \\
 \textbf{2005 RO43}      	&28.817   &0.520   &13.832    &43.802   &35.5     &194 \\
 2007 JK43      		&46.604   &0.494   &23.559    &69.649   &44.8     &176 \\
 2004 PY117     		&40.050   &0.284   &28.679    &51.420   &23.4     &176 \\
 2011 UH413     		&38.182   &0.552   &17.102    &59.262   &46.6     &176 \\
 2009 JE19      		&43.495   &0.438   &24.425    &62.565   &22.0     &168 \\
  \textbf{2011 WG157}     	&29.927   &0.031   &28.995    &30.859   &22.3     &168 \\
 2011 KT19      		&35.581   &0.331   &23.797    &47.366   &110.1    &161 \\
  \textbf{2010 FB49}      	&22.613   &0.191   &18.292    &26.934   &24.3     &154 \\
\hline
\multicolumn{7}{l}{Boldface - small bodies with  $a\leqslant30$ au (Centaurs)}
\end{tabular}
 \end{table*}

\section{Summary, Conclusions and prospects for detecting rings around other Centaurs}
\label{sec_final}
The problem of ringed small bodies is quite recent. The first confirmed case of this problem was the discovery of the rings around the Centaur Chariklo \citep{braga1,berard}. Later, it was proposed that the Centaur (2060) Chiron may also have a ring \citep{ortiz}.
More recently, the discovery that the dwarf planet (136108) Haumea also has a ring was announced \citep{ortiz-haumea}.

In \cite{araujo2016}, we showed that the rings of Chariklo may experience a propitious environment for the existence of rings within the Centaur region.
The possible existence of a second ringed Centaur, which has not yet been confirmed, raises the question of whether the results found for Chariklo`s rings are also valid for Chiron`s ring or for the other Centaurs in general.

Given this scenario, we present a similar study where the method adopted in our previous study for Chariklo is applied for Chiron. We considered a sample of clones of Chiron, and the orbits of those clones were numerically integrated over time while considering the giant planets and the Sun.

Using the tidal disruption radius $r_{td}$ as the limit that defines a disruptive encounter, we computed the number of close encounters suffered by Chiron and its ring with the giant planets within this distance and along its half-life ($0.36$ Myr). The same procedure was applied to Chariklo by using the simulation data of our previous work and considering the half-life of Chariklo ($7$ Myr). 

We found that Chiron is more likely to lose its rings than Chariklo. We estimated that the probability of Chiron losing its rings is $\approx 5.7$ times higher than that of Chariklo.

The distinct masses and ring orbital radii of these systems lead to a relatively small difference in the number of disruptive encounters that each system would suffer when in the same orbit. On average, the number of disruptive encounters of Chiron was increased by a factor of $1.4$ when compared to that of Chariklo.

The analysis of the orbital evolutions of Chiron and of Chariklo showed that Chiron is more likely to remain in a relatively low-inclination and high-eccentricity orbit with a pericentre of less than $10$ au, meaning that Chiron is more likely to remain in an orbit that crosses the orbits of Saturn and Jupiter. In fact, we see that more than $30\%$ of the clones of Chiron spent most of their time in orbits with pericentre distances inside the orbit of Jupiter.

Conversely, Chariklo is more likely to remain in a relatively high-inclination and low-eccentricity orbit with a pericentre greater than $10$ au. Thus, Chariklo is more likely to remain in an orbit that does not cross the orbit of Saturn. This clearly represents a lower chance of Chariklo suffering disruptive close encounters with the giant planets than that for Chiron.

Based on these analyses, we found that the orbital evolution of Chiron is much less favourable, by a factor of $\approx 4.5$, for the existence of rings than the orbital evolution of Chariklo.

Therefore, we found that the higher probability of the disruption of Chiron's ring than that for Chariklo, given as a factor of $\approx6$, is mainly a consequence of how the heliocentric orbits of these systems evolve and is not due to the configurations of the ringed systems themselves.

Overall, we can state that a ring of a Centaur has a higher probability of survival in the giant planets region if the central body is in a Chariklo-like orbit, i.e., in a relatively high-inclination and low-eccentricity orbit. In addition, more massive bodies also provide better conditions for the survival of a ring since more massive bodies result in smaller tidal disruption radii, meaning that the system has to come closer to a planet to be disrupted.

For observational purposes, our results indicate that if the goal is to observe Centaurs with higher probabilities of preserving possible rings, then bigger bodies with high inclinations and low eccentricities should be prioritized. 

By applying these criteria to the list given by Johnston, W.R.\footnote{http://www.johnstonsarchive.net/astro/tnoslist.html}, we selected a set of small non-resonant bodies with diameters greater than the value assumed here for Chiron in orbits with distances of the pericentre and distances of the apocentre less than $30$ au and $100$ au, respectively, and with orbital inclinations greater than $20^{\circ}$. We found 17 bodies that meet these conditions. They are listed in Table 4 based on their sizes in descending order. The bodies with $a\leqslant30$ au are classified as Centaurs and are listed in boldface type. Note that Chariklo appears here as the Centaur with the most promising conditions for maintaining rings. 

Haumea, whose orbital elements$^2$ are $a=43.2$ au, $e=0.2$, $i=28.2^{\circ}$, was not included in Table 4 because the distance of pericentre of its orbit is $R_p\approx35$ au, i.e., Haumea is currently a TNO. Nevertheless, in agreement with our results, we see that the orbital inclination and the size of Haumea, in fact, make this body favourable to maintain a ring.

\section{Acknowledgements}
This work was funded by CAPES, CNPq Procs. 305737/2015-5 and 312813/2013-9, and by 
FAPESP Procs. 2016/24561-0 and 2011/08171-3. 
This support is gratefully acknowledged. 
The authors are also grateful to the PROPE-UNESP for the financial support.

\renewcommand{\refname}{REFERENCES}

\label{lastpage}

\end{document}